# On the theory of proton solid echo in polymer melts

N. Fatkullin,[1] A. Lozovoi[2], C. Mattea,[2] S. Stapf[2]
[1]Institute of Physics, Kazan Federal University, Kazan, 420008, Tatarstan, Russia,
[2]Technische Universität Ilmenau, Dept. Technical Physics II, 98684 Ilmenau, Germany.

**Abstract**

Based on a modified Anderson-Weiss approximation (N. Fatkullin, A. Gubaidullin, C. Mattea, S.Stapf, J. Chem. Phys. 137 (2012), 224907) an improved theory of proton spin solid echo in polymer melts is formulated, taking into account contribution from intermolecular magnetic dipole-dipole interactions. The solid echo build-up function defined by the relation $I^{SE}(2\tau;\tau) \equiv \dfrac{S_1(2\tau;\tau) + S_2(2\tau;\tau) - S_3(2\tau;\tau)}{S_1(2\tau;\tau) + S_2(2\tau;\tau)}$, where $S_1(2\tau;\tau)$, $S_2(2\tau;\tau)$ and $S_3(2\tau;\tau)$ are the respective signals arising from ($\hat{P}_x^{\pi/2} - \tau - \hat{P}_y^{\pi/2}$),($\hat{P}_x^{-\pi/2} - \tau - \hat{P}_x^{-\pi/2}$) and ($\hat{P}_x^{-\pi/2} - \tau - \hat{P}_x^{\pi}$) spin echo experiments, where $\hat{P}_\alpha^\theta$ is an operator rotating the spin system on the angle $\theta$ relatively axis $\alpha$, is investigated. It is shown that the intermolecular part of this function at short times $t < \tau_{fl}$, where $\tau_{fl}$ is a characteristic time for flip-flop transitions between proton spins, contains information about the relative mean squared displacements of polymer segments at different macromolecules, opening up a new opportunity for obtaining information about polymer dynamics in the millisecond regime.

## 1. Introduction.

Proton NMR is a traditional and widely used method for experimental investigations of structure and dynamics in different fields of condensed matter in general, and polymer physics in particular [1-13]. In most general terms, this method is based on affecting the equilibrium spin system of an investigated sample by particular sequences of radio frequency (RF) pulses and studying their response, i.e. the evolution of the spin system, or spin kinetics, with time. The dependence of this observed kinetics on experimentally controlled parameters such as evolution time, resonance frequency, architecture of RF pulses contains important information about the spatial translations of spin-bearing hydrogen nuclei of the investigated sample.

The proton spin dynamics is mainly controlled by magnetic dipole-dipole interactions between different protons, which can be classified as intramolecular and intermolecular. For the case of polymer systems, intramolecular magnetic dipole-dipole interactions can further be divided into intrasegment interactions, i.e. interactions between spins from the same Kuhn segment, and segment-segment interactions, i.e. between spins from different Kuhn segments of the same macromolecule. During many decades it was postulated that the main contribution is a result of intramolecular magnetic dipole-dipole interactions between protons belonging to the same polymer segment, see for example, [4,5, 8,10,16] and literature cited therein.

In [17-19] it was theoretically argued and experimentally shown, that this postulate is very far from reality at least for proton spin-lattice relaxation in polymer melts. Moreover, the separation of intermolecular and intramolecular contributions to the spin-lattice relaxation rate and their frequency dispersion allow one to extract the time-dependence of the relative mean squared displacement of polymer segments of different polymer chains. Then essential development both experimentally and theoretically of discussed method covering the time interval $10^{-9} < t < 10^{-3} s$ for the study of macromolecular dynamics in polymer melts was described in [20-24].

The timescale of milliseconds can also be studied by other proton NMR methods: Double Quantum Resonance (DQ resonance) [25-36], Free Induction Decay (FID), or by the equivalent Hahn Echo, and Solid Echo [16, 37-45]. The theory of proton FID, Hahn Echo and DQ resonance in polymer melts with taking into account magnetic dipole-dipole interactions was constricted in papers [46,47], but the theory of proton Solid Echo in polymer melts has, to the best of our knowledge, only been formulated on a level of consideration exclusively intrasegment proton magnetic dipole-dipole interactions. Advancing this simplified approach in the theory of the proton Solid Echo in polymer melts is a main goal of this paper.

Actually we will discuss, as has been done in [16], the superposition of two variants of Solid Echo, which differ from each other in the second RF pulse, and the Hahn Echo, i.e. a superposition of signals [$S_1$-$S_2$-$S_3$], where $S_1$, $S_2$ and $S_3$ are the respective signals arising from ($\hat{P}_x^{\pi/2}-\tau-\hat{P}_y^{\pi/2}$),($\hat{P}_x^{-\pi/2}-\tau-\hat{P}_x^{-\pi/2}$) and ($\hat{P}_x^{-\pi/2}-\tau-\hat{P}_x^{\pi}$) spin echo experiments where $\hat{P}_\alpha^\theta$ is an operator rotating the spin system on the angle $\theta$ relatively axis $\alpha$ (the signs of rotation angles were chosen in the way that all initial values of signals following these sequences are positive). From these signals it is possible to construct the function $I^{SE}(2\tau;\tau) \equiv \dfrac{S_1(2\tau;\tau)+S_2(2\tau;\tau)-S_3(2\tau;\tau)}{S_1(2\tau;\tau)+S_2(2\tau;\tau)}$, where $\tau$ is the time interval between two RF pulses and $2\tau$ is the time at which the corresponding echo is being observed. This function, as will be shown later, at times $\tau < T_2$, where $T_2$ is the proton spin-spin relaxation time caused by the magnetic dipole-dipole interactions, contains contributions both from inter- and intramolecular interactions. In high molecular mass polymer melts at times shorter than the polymer melt terminal relaxation time, due to anomalous character of polymer diffusion, the intermolecular contribution is determined by relative mean squared displacements of polymer segments from different polymer chains.

## 2. General theoretical consideration.

Following the argumentation in a recent paper [46] we will discuss a system described by the following Hamiltonian:

$$\hat{H} = \hat{H}_s + \hat{H}_L + \hat{H}_{dd}^{\sec}, \qquad (1)$$

where $\hat{H}_s = \sum_k \hbar\omega_0 \hat{I}_k^z$ is the spin Hamiltonian of the Zeeman interaction describing the interaction of proton spins with the external magnetic field, $\hbar$ is the Planck constant, $\omega_0$ is the resonance frequency, $\hat{H}_L$ is the Hamiltonian of the lattice degrees of freedom, describing motions of macromolecules in space.

The secular part of the Hamiltonian of the magnetic dipole-dipole interactions, giving, at high resonance frequencies, the main contribution to the transverse relaxation, can be written in the standard way:

$$\hat{H}_{dd}^{\sec} = \frac{1}{2} \sum_{k;l} A_{kl} \left( \hat{I}_k^z \hat{I}_l^z - \frac{1}{4} \left( \hat{I}_k^+ \hat{I}_l^- + \hat{I}_k^- \hat{I}_l^+ \right) \right), \tag{2}$$

where for $i \neq j$

$$A_{kl} = \frac{\gamma^2 \hbar^2}{r_{kl}^3} \left( 1 - 3\cos^2(\theta_{kl}) \right), \tag{3}$$

$\gamma$ is the gyromagnetic ratio of proton spins, $k$ and $l$ are indexes enumerating spins, $r_{kl}$ is the distance between spins with numbers $k$ and $l$, $\theta_{kl}$ is the angle between the direction of the external magnetic field and vector $\vec{r}_{kl}$ connecting the two considered spins, $\hat{I}_k^z, \hat{I}_k^+ = \hat{I}_k^x + i\hat{I}_k^y, \hat{I}_k^- = \hat{I}_k^x - i\hat{I}_k^y$ are z-component, raising and lowering spin operators, respectively, of the spin with number $k$, $A_{ii} = 0$.

The enumerating indices in our case have a complex character, i.e. they implicitly include additional sub-indices corresponding to the number of the macromolecule, segment in the macromolecule and the spin inside the segment.

The secular part of the spin-spin interaction Hamiltonian, to which we include, for the sake of generality, apart from the magnetic dipole-dipole interaction also a possible scalar exchange interaction, can be represented as the following:

$$\hat{H}_{dd}^{\sec} = \frac{3}{4} \sum_{k;l} A_{kl} \hat{I}_k^z \hat{I}_l^z - \frac{1}{4} \sum_{k;l} \tilde{A}_{kl} \hat{\vec{I}}_k \cdot \hat{\vec{I}}_l, \tag{4}$$

where $\hat{\vec{I}}_k \cdot \hat{\vec{I}}_l = \hat{I}_k^x \hat{I}_l^x + \hat{I}_k^y \hat{I}_l^y + \hat{I}_k^z \hat{I}_l^z$,

$\tilde{A}_{kl} = A_{kl} - 2J_{kl}$, $J_{kl}$ is the constant of an exchange interaction between spins with numbers $k$ and $l$. (Note that this is a correction of a misprint in the corresponding expression of [48] and that the numerical coefficient has been changed accordingly.)

The separation (4) is convenient in the theory of magnetic resonance, because it is well known that the scalar part of the right part of the expression (4) does not give a contribution to the second moment of magnetization (see, for example [1-3]), and therefore to the evolution of the spin system at times $t < T_2$, where $T_2$ is the spin-spin relaxation time, i.e. characteristic decay time for Free Induction Decay of a spin system governed by the Hamiltonian (1). Another point which favors the approach of the separation (4) is the fact that many steps of the evolution of the spin system can be calculated exactly if the scalar part of the Hamiltonian (4) is neglected. In addition, the use of the standard Anderson–Weiss approximation gives very reasonable results in many cases. The Anderson-Weiss approximation is nothing more than the magnetic resonance version of the second cumulant approximation in the general statistical mechanics, which ignores the scalar part of the Hamiltonian (4). This part is responsible for interspin flip-flop processes, which create the phenomenon called spin-diffusion. In our recent paper [46] we suggested the modified Anderson-Weiss approximation for Free Induction Decay and Hahn Echo signal and took into account the existence of the scalar part in the Hamiltonian and therefore described the effects of the spin-diffusion. It was shown that scalar part of the Hamiltonian (4) influences the relative mean-squared displacements calculated from intermolecular contribution to FID at times $t \leq 2T_2$ by less than 10%, a value that is fully acceptable considering typical experimental accuracy.

At the initial moment of time the state of the total system, i.e. spins + lattice, is described by the equilibrium density matrix

$$\hat{\rho}^{eq} = \frac{1}{Z}\exp\{-\beta\hat{H}\} \cong \hat{\rho}_s^{eq}\hat{\rho}_L^{eq} = \frac{1}{Z_s}\exp\{-\beta\hat{H}_s\}\hat{\rho}_L^{eq}$$
$$\cong \frac{1}{Z_s}\left(I - \beta\hbar\omega_0\hat{I}_z\right)\hat{\rho}_L^{eq}$$
(5)

where $\beta = (k_B T)^{-1}$ is the inverse temperature of the system, $k_B$ is the Boltzmann constant, $\hat{\rho}_L^{eq}$ is the equilibrium density matrix of the lattice, $Z_s \cong (2I+1)^{N_s}$ is the statistical sum of the spin system in the high temperature approximation, i.e. $\beta\hbar\omega_0 \ll 1$ which is valid with a high accuracy at any temperature above tens millikelvin, $N_s$ is the total number of spins in the system with the resonance frequency $\omega_0$, $\hat{I}_z = \sum_k \hat{I}_k^z$.

### 2.1. Calculation of the response signal $S_1$.

As was already mentioned in the Introduction section, this signal is the response of the spin system to the pulse sequence $\hat{P}_x^{\pi/2} - \tau - \hat{P}_y^{\pi/2}$. After application of the first RF pulse rotating the spin system by an angle $\pi/2$ about the X axis the equilibrium density matrix becomes:

$$\hat{\rho}_0 = \hat{P}_x^{\pi/2} \hat{\rho}^{eq} \equiv \exp\left\{-i\frac{\pi}{2}\hat{I}_x\right\} \hat{\rho}^{eq} \exp\left\{i\frac{\pi}{2}\hat{I}_x\right\} \cong \frac{1}{Z_s}\left(I + \beta\hbar\omega_0 \hat{I}_y\right)\hat{\rho}_L^{eq}. \tag{6}$$

After this the system follows the free evolution determined by the Hamiltonian (1) and the density matrix of the whole system at the time moment $t$ is equal to

$$\hat{\rho}(t) = \hat{S}(t)\hat{\rho}_0 = \exp\{-it\hat{L}_H\}\hat{\rho}_0, \tag{7}$$

where, for the purpose of abbreviation, the Liouville space formalism is used (see, for example [3], i.e.

$\hat{S}(t)$ is the superoperator of evolution caused by the Hamiltonian $\hat{H}$, which, by the definition, is acting in accordance with the following rule:

$$\hat{S}(t)\hat{\rho}_0 \equiv \exp\{-i\hat{H}t\}\hat{\rho}_0 \exp\{i\hat{H}t\}, \tag{8}$$

and $\hat{L}_H$ is the Liouville operator defined by the relation

$$\hat{L}_H\hat{\rho} \equiv \frac{1}{\hbar}\left[\hat{H};\hat{\rho}\right]. \tag{9}$$

Subsequently at time $t = \tau$, the second RF pulse $\hat{P}_y^{\pi/2}$ acts on the spin system and rotates the spin system on the angle $\pi/2$ about Y axis. Note that the final result will not change if the second RF pulse is $\hat{P}_y^{-\frac{\pi}{2}}$, i.e. it rotates the spin system on the angle $-\pi/2$ relatively axis Y. If one uses $\hat{P}_x^{-\pi/2}$ as the first RF pulse, then the final result will change sign, but will still have the same value as in the case we are discussing now.

Consider the situation when the experimentally measurable quantity is the y component of the total spin of the system, and can be calculated with the help of the statistical operator at time $t$ by the following standard relation:

$$S_1(t) \equiv \left\langle \hat{I}_y(t)\right\rangle_1 = Tr\left(\hat{I}_y\hat{\rho}(t)\right) \cong \frac{\beta\hbar\omega_0}{(2I+1)^{N_s}} Tr\left(\hat{I}_y\hat{S}(t-\tau)\hat{P}_y^{\pi/2}\hat{S}(\tau)\hat{I}_y\hat{\rho}_L^{eq}\right), \tag{10}$$

where $Tr(...)$ is the trace operation over all spin and lattice variables, and $N_s$ is the number of all resonant proton spins in system, $\hat{I}_y = \sum_k \hat{I}_k^y$.

Note that for simplicity we are doing all the calculations in the laboratory frame, while in the real experiment RF pulses are being applied in the so called rotating frame. At time moment $t = 2\tau$ when echo is being observed, results are the same both in laboratory and in rotating frames.

Within the accuracy of the high-temperature approximation, i.e. to within about $10^{-5}$ at room temperature, the expression (10) is exact. Further evaluation of it demands approximations due to the presence of the multi-particle interaction term $\hat{H}_{dd}^{sec}$ given by expression (2) in the total Hamiltonian (1). We will use further on the modified Anderson-Weiss approximation, details of which can be found in [46].

A first step of this approximation is traditional and is based on the transition to the so-called interaction, or Dirac, representation. The main difference to usual standard schemes contains in choosing the so-called "zero Hamiltonian", which in our case includes the scalar part of the Hamiltonian (1), defined as:

$$\hat{H}_0 = \hat{H}_s + \hat{H}_L - \frac{1}{4}\sum_{k;l} \tilde{A}_{kl} \hat{\vec{I}}_k \cdot \hat{\vec{I}}_l \ . \tag{11}$$

Therefore in our case the role of perturbation is assumed by the following Hamiltonian:

$$\hat{H}_{dd}^{sec;zz} = \frac{3}{4}\sum_{k;l} A_{kl} \hat{I}_k^z \hat{I}_l^z \ . \tag{12}$$

Expression (10) can be rewritten as:

$$S_1(t) = \frac{\beta\hbar\omega_0}{\left(2I+1\right)^{N_s}} Tr\left( \hat{I}_y \hat{S}_0(t) \hat{\tilde{S}}_{dd}^{sec}(t-\tau;\tau) \hat{S}_0^{-1}(\tau) \hat{P}_y^{\pi/2} \hat{S}_0(\tau) \hat{\tilde{S}}_{dd}^{sec}(\tau;0) \hat{I}_y \hat{\rho}_L^{eq} \right), \tag{13}$$

where $\quad \hat{S}_0(t) = \exp\{-it\hat{L}_0\} \tag{14}$

is the superoperator of evolution created by the Hamiltonian (11) and

$$\hat{\tilde{S}}_{dd}^{sec}(t_2;t_1) = \hat{T}\exp\left\{-i\int_{t_1}^{t_2} \hat{\tilde{L}}_{dd}^{sec;zz}(t')dt'\right\} \ . \tag{15}$$

is the superoperator of evolution created by the Hamiltonian (12) in the interaction representation, where $\hat{T}$ means the usual Dyson time ordering operator.

Then, for the calculation with superoperator (15), one can use the standard quantum statistical perturbation theory to (15) cutting the series decomposition on terms having second order of magnitude with respect to $\hat{\hat{L}}_{dd}^{\mathrm{sec};zz}(t)$. Then contributions of higher orders of magnitude can be approximately recovered using the second cumulant, i.e. the Anderson-Weiss, approximation for calculating the spin echo signal. For realization of this procedure it is necessary to be able to calculate the time evolution of operators having the structure:

$$\hat{S}_0^*(t)\hat{I}_k^z\hat{I}_l^z = \exp\left\{i\frac{\hat{H}_0}{\hbar}t\right\}\hat{I}_k^z\hat{I}_l^z \exp\left\{-i\frac{\hat{H}_0}{\hbar}t\right\} . \tag{16}$$

In our case the zero Hamiltonian includes the scalar part of the spin-spin interactions and due to that the right part of expression (16) can not be calculated exactly. An approximation suggested in [46] consists of the right-hand side of operators having a structure similar to (16) by it's projection in the sense of Zwanzig-Mori, see for example [2]:

$$\hat{S}_0^*(t)\hat{I}_k^z\hat{I}_l^z \approx \hat{P}_{kl}^{zz}\hat{S}_0^*(t)\hat{I}_k^z\hat{I}_l^z \equiv \hat{I}_k^z\hat{I}_l^z \frac{Tr\left(\hat{I}_k^z\hat{I}_l^z\hat{S}_0^*(t)\hat{I}_k^z\hat{I}_l^z\hat{\rho}_L^{eq}\right)}{Tr_s\left(\hat{I}_k^z\hat{I}_l^z\right)^2} =$$
$$= \hat{I}_k^z\hat{I}_l^z P_{kl}^{fl}(t) \tag{17}$$

where $Tr_s(...)$ is the trace operation over all the spin variables.

Note that experimentally measurable quantities are time dependent correlation functions having structures similar to (13). They have translational invariant symmetry relatively displacements of initial moment of time, i.e. they do not depend on the choice of the initial moment of time. This symmetry is exact. The approximation (17) obviously does not possess this symmetry. This means, that (17) should be applied with additional instructions to keep discussed symmetry. Firstly, many time-dependent correlation functions using exact symmetry properties should be rewritten into some "normal form". This means, that using invariance relatively time displacements property they should be rewritten to form, in which the time argument of spin operators like $\hat{S}_0^*(t_i)\hat{I}_k^z\hat{I}_l^z$ with the earliest moment of time is zero.

Then the approximation (17) should be applied to these time-dependent correlation functions written in the "normal form". For example, let us discuss how to calculate a quantity, which has a structure $J(t_2;t_1) = Tr\left(\left(\hat{S}_0^*(t_i)\hat{I}_k^z\hat{I}_l^z\right)\hat{B}\left(\hat{S}_0^*(t_i)\hat{I}_k^z\hat{I}_l^z\right)\hat{\rho}_L^{eq}\right)$, where $\hat{B}$ is a time independent operator. Using translational invariance we can rewrite it into a normal form, which is $J(t_2;t_1) = Tr\left(\left(\hat{S}_0^*(t_2-t_1)\hat{I}_k^z\hat{I}_l^z\right)\hat{B}\hat{I}_k^z\hat{I}_l^z\hat{\rho}_L^{eq}\right)$ Now we can apply the approximation (17) and obtain $J(t_2;t_1) = P_{kl}^{fl}(t_2-t_1)Tr\left(\hat{I}_k^z\hat{I}_l^z\hat{B}\hat{I}_k^z\hat{I}_l^z\hat{\rho}_L^{eq}\right)$.

The quantity $P_{kl}^{fl}(t)$ can be considered as the probability for a given pair of spins with numbers $k$ and $l$ not to participate in flip-flop processes in the time interval $t$. For protons having spin $I = \frac{1}{2}$ mutual flip-flop transitions between spins with numbers $k$ and $l$ do not give a contribution to the probability $P_{kl}^{fl}(t)$. For discussing this probability the following expression was derived using standard Anderson-Weiss approximation (see [46]):

$$P_{kl}^{fl}(t) = \exp\left\{-\int_0^t d\tau(t-\tau)\frac{I(I+1)}{6\hbar^2}\sum_m\left(\left\langle\tilde{A}_{km}(\tau)\tilde{A}_{km}(0)\right\rangle_{eq} + \left\langle\tilde{A}_{lm}(\tau)\tilde{A}_{lm}(0)\right\rangle_{eq}\right)\right\}, \qquad (18)$$

for proton spins $I = \frac{1}{2}$.

Expression (13) for the signal $S_1$ can be rewritten in the form:

$$S_1(t) = \frac{\beta\hbar\omega_0}{(2I+1)^{N_s}}Tr\left(\left(\hat{S}_0(\tau)\left(\hat{\tilde{S}}_{dd}^{sec}(t-\tau;\tau)\right)^{-1}\hat{S}_0^{-1}(t)\hat{I}_y\right)\hat{P}_y^{\pi/2}\hat{S}_0(\tau)\hat{\tilde{S}}_{dd}^{sec}(\tau;0)\hat{I}_y\hat{\rho}_L^{eq}\right). \qquad (19)$$

Employing the approximation (17) the action of evolution superoperators on the spin variables $\hat{S}_0(\tau)\left(\hat{\tilde{S}}_{dd}^{sec}(t-\tau;\tau)\right)^{-1}\hat{S}_0^{-1}(t)\hat{I}_y$ and $\hat{S}_0(\tau)\hat{\tilde{S}}_{dd}^{sec}(\tau;0)\hat{I}_y$ can be calculated exactly. Then, using properties of spin $I = \frac{1}{2}$, symmetry arguments like the isotropy of system, considering motions of lattice variables classically, after somewhat bulky quantum statistical calculations which have been described in details in [46], one obtains the following result:

$$S_1(t;\tau) =$$

$$= \frac{\beta\hbar\omega_0}{4} \sum_k \left\langle \begin{array}{c} \left(\cos\left(\tilde{\varphi}_k(t-\tau)\right)\cos\left(\varphi_k(\tau)\right)\right) \times \\ \left(\sum_s \cos\left(\frac{1}{2}\tilde{\varphi}^d_{ks}(t-\tau) - \frac{1}{2}\varphi^d_{ks}(t-\tau)\right) \prod^*_{m;l} \cos\left(\frac{1}{2}\tilde{\varphi}^d_{km}(t-\tau)\right) \cos\left(\frac{1}{2}\varphi^d_{kl}(\tau)\right)\right) \end{array} \right\rangle, \quad (20)$$

where

$$\varphi^d_{kl}(\tau) = \frac{3\gamma^2\hbar}{2} \int_0^\tau dt_1 \frac{1-3\cos^2\left(\theta_{kl}(t_1)\right)}{r^3_{kl}(t_1)} P^{fl}_{kl}(t_1), \quad (21)$$

$$\tilde{\varphi}^d_{km}(t-\tau) = \frac{3\gamma^2\hbar}{2} \int_\tau^t dt_1 \frac{1-3\cos^2\left(\theta_{km}(t_1)\right)}{r^3_{km}(t_1)} P^{fl}_{km}(t_1), \quad (22)$$

the phases $\varphi_k(\tau) = \omega_k \cdot \tau$ and $\tilde{\varphi}_k(t-\tau) = \omega_k \cdot (t-\tau)$ are connected with either chemical shift differences of different protons or different Larmor frequencies caused by external magnetic field gradient, the latter being assumed small enough for neglecting diffusion effects, and $\prod^*_{m;l}(...)$ means, that inside bracket $k,s \neq m,l$.. The bracket $\langle...\rangle$ denotes, as usual, the equilibrium averaging over lattice variables. The quantities $\varphi^d_{kl}(\tau)$ and $\tilde{\varphi}^d_{km}(t-\tau)$ are related to rotations of proton spins in local dipolar fields after the first and the second RF pulses, respectively, and contain information about polymer segments dynamics through time dependence of the factors $\frac{1-3\cos^2\left(\theta_{kl}(t_1)\right)}{r^3_{kl}(t_1)}$ inside integrals at the right-hand side of expressions (21) and (22). Due to the factors $\left\langle\cos\left(\tilde{\varphi}_k(t-\tau)\right)\cos\left(\varphi_k(\tau)\right)\right\rangle$ the signal $S_1(t;\tau)$ has a maximum, i.e. an echo, at time $t = 2\tau$.

### 2.2. Calculation of the response signals $S_2$ and $S_3$.

In the case of $S_2$ we have the response of the spin system to the pulse sequence $\hat{P}^{-\pi/2}_x - \tau - \hat{P}^{-\pi/2}_x$, i.e. the situation when both RF pulses rotate the spin system by an angle $-\pi/2$ about the X. (If the sequence $\hat{P}^{\pi/2}_x - \tau - \hat{P}^{\pi/2}_x$ is used, the result will have the opposite sign.) The experimentally observed quantity is the y component of the total spin of the system. It can be calculated in accordance with the following expression:

$$S_2(t) \equiv \left\langle \hat{I}_y(t) \right\rangle_2 = Tr\left(\hat{I}_y \hat{\rho}(t)\right) \cong -\frac{\beta \hbar \omega_0}{(2I+1)^{N_s}} Tr\left(\hat{I}_y \hat{S}(t-\tau) \hat{P}_x^{-\pi/2} \hat{S}(\tau) \hat{I}_y \hat{\rho}_L^{eq}\right). \tag{23}$$

Calculation of the expression (23) can be made analogously to the expression (13). Firstly, it is possible to rewrite it as:

$$S_2(t) = -\frac{\beta \hbar \omega_0}{(2I+1)^{N_s}} Tr\left(\left[\hat{S}_0(\tau)\left(\hat{\tilde{S}}_{dd}^{sec}(t-\tau;\tau)\right)^{-1} \hat{S}_0^{-1}(t) \hat{I}_y \right] \hat{P}_x^{-\pi/2} \hat{S}_0(\tau) \hat{\tilde{S}}_{dd}^{sec}(\tau;0) \hat{I}_y \hat{\rho}_L^{eq}\right). \tag{24}$$

Then after using approximation (17), taking into account symmetry arguments and special algebraic properties of spin operators for spin $I = \frac{1}{2}$, expression (24) can be transformed to the following:

$$S_2(t;\tau) =$$
$$= \frac{\beta \hbar \omega_0}{4} \sum_k \left\langle \begin{array}{l} \sin\left(\tilde{\varphi}_k(t-\tau)\right) \sin\left(\varphi_k(\tau)\right) \times \\ \left(\sum_s \cos\left(\frac{1}{2}\tilde{\varphi}_{ks}^d(t-\tau) - \frac{1}{2}\varphi_{ks}^d(t-\tau)\right)\right) \prod_{m;l}^* \cos\left(\frac{1}{2}\tilde{\varphi}_{km}^d(t-\tau)\right) \cos\left(\frac{1}{2}\varphi_{kl}^d(\tau)\right) \end{array} \right\rangle. \tag{25}$$

This expression is very similar to the expression (20) for the signal $S_1$ except for a difference of the factors, which are responsible for the echo signal at time $t = 2\tau$: expression (25) contains factors $\sin\left(\tilde{\varphi}_k(t-\tau)\right)\sin\left(\varphi_k(\tau)\right)$ instead of $\cos\left(\tilde{\varphi}_k(t-\tau)\right)\cos\left(\varphi_k(\tau)\right)$ in expression (20).

The calculation of the signal $S_3$, i.e. response of the spin system to the pulse sequence $\hat{P}_x^{-\pi/2} - \tau - \hat{P}_x^{\pi}$, which is frequently called Hahn Echo, can be calculated analogously to the signals $S_1$ and $S_2$ and the result is the following:

$$S_3(t;\tau) = \frac{\beta \hbar \omega_0}{4} \sum_k \left\langle \cos\left(\tilde{\varphi}_k(t-\tau) - \varphi_k(\tau)\right) \prod_m \cos\left(\frac{1}{2}\left(\tilde{\varphi}_{km}^d(t-\tau) + \varphi_{km}^d(\tau)\right)\right) \right\rangle. \tag{26}$$

The most important difference of the Hahn echo, the signal $S_3$, to the two variants of the solid echo, signal $S_1$, which is usually named the solid echo, and signal $S_2$, occurs in the factor $\cos\left(\frac{1}{2}\left(\tilde{\varphi}_{km}^d(t-\tau) + \varphi_{km}^d(\tau)\right)\right)$, where it can be seen that the influence of local dipolar fields on the spin evolution is additive, i.e. the expression contains the sum of phases $\frac{1}{2}\left(\tilde{\varphi}_{km}^{d}(t-\tau)+\varphi_{km}^{d}(\tau)\right)$, the factor $\frac{1}{2}$ is reflecting the fact that one is dealing with spins $I=\frac{1}{2}$. At time $t=2\tau$ the expression (24) is equivalent to the expression for FID derived in paper [46].

### 2.3. The solid echo build up function $I^{SE}(2\tau;\tau)$.

Consider the sum of two echo signals $S_1$ and $S_2$:

$$S_{12}(t;\tau) \equiv S_{1}(t;\tau)+S_{2}(t;\tau) = \frac{\beta\hbar\omega_{0}}{4}\sum_{k}\left\langle\begin{array}{l}\cos\left(\tilde{\varphi}_{k}(t-\tau)-\varphi_{k}(\tau)\right)\times\\ \sum_{s}\cos\left(\frac{1}{2}\tilde{\varphi}_{ks}^{d}(t-\tau)-\frac{1}{2}\varphi_{ks}^{d}(t-\tau)\right)\prod_{m;l}^{*}\cos\left(\frac{1}{2}\tilde{\varphi}_{km}^{d}(t-\tau)\right)\cos\left(\frac{1}{2}\varphi_{kl}^{d}(\tau)\right)\end{array}\right\rangle \quad (27)$$

This sum has the same echo forming factors $\cos\left(\tilde{\varphi}_{k}(t-\tau)-\varphi_{k}(\tau)\right)$ as the Hahn echo (26).

All the discussed signals have a maximal value at $t=2\tau$ and it is useful to define the following normalized $I^{SE}(2\tau;\tau)$ function which reflects the difference between solid echoes and Hahn echo and we will name it as the solid echo build up function:

$$I^{SE}(2\tau;\tau) \equiv \frac{S_{12}(2\tau;\tau)-S_{3}(2\tau;\tau)}{S_{12}(2\tau;\tau)}. \quad (28)$$

At $t=2\tau$ one has $\cos\left(\tilde{\varphi}_{k}(t-\tau)-\varphi_{k}(\tau)\right)\approx 1$ and, in particular for polymer melts with large molecular masses, the dependence of the particular contributions inside the sums in expressions (26) and (27) on the spin number $k$ is very weak, i.e. they are equal to each other. Therefore one can see from expressions (26), (27) and (28):

$$I^{SE}(2\tau;\tau)=1-\frac{\sum_{k}\prod_{m}\left\langle\cos\left(\frac{1}{2}\left(\tilde{\varphi}_{km}^{d}(t-\tau)+\varphi_{km}^{d}(\tau)\right)\right)\right\rangle}{\sum_{k}\left\langle\left(\sum_{s}\cos\left(\frac{1}{2}\tilde{\varphi}_{ks}^{d}(t-\tau)-\frac{1}{2}\varphi_{ks}^{d}(t-\tau)\right)\right)\prod_{m;l}^{*}\cos\left(\frac{1}{2}\tilde{\varphi}_{kl}^{d}(t-\tau)\right)\cos\left(\frac{1}{2}\varphi_{km}^{d}(\tau)\right)\right\rangle}. \quad (29)$$

Note that for a two-spin system the introduced function $I^{SE}(2\tau;\tau)$ is analogous to the $\beta(2\tau;\tau)$ function discussed in paper [16], see also close approaches in [4] and literature sited therein. A many-spin generalization of the $\beta(2\tau;\tau)$ function can be obtained from the expression (28), if one replaces $S_{12}(2\tau;\tau)$ in the denominator with its initial value $S_{12}(2\tau=0;\tau=0)$, which is actually not easy to determine experimentally.

Then, employing the following approximation for the cosine function in (29):

$$\cos(x) = 1 - \frac{1}{2}x^2 + ... \cong \exp\left\{-\frac{1}{2}x^2\right\}, \tag{30}$$

Expression (29) can be evaluated to the form:

$$I^{SE}(2\tau;\tau) = 1 - \frac{\sum_k \exp\left\{-\frac{1}{8}\sum_m \left[\left\langle\left(\tilde{\varphi}_{km}^d(\tau)\right)^2\right\rangle + 2\left\langle\tilde{\varphi}_{km}^d(\tau)\cdot\varphi_{km}^d(\tau)\right\rangle + \left\langle\left(\varphi_{km}^d(\tau)\right)^2\right\rangle\right]\right\}}{\sum_k \exp\left\{-\frac{1}{8}\sum_m \left[\left\langle\left(\tilde{\varphi}_{km}^d(\tau)\right)^2\right\rangle - 2\left\langle\tilde{\varphi}_{km}^d(\tau)\cdot\varphi_{km}^d(\tau)\right\rangle + \left\langle\left(\varphi_{km}^d(\tau)\right)^2\right\rangle\right]\right\}}. \tag{31}$$

For the situations, when all the spins are equivalent, i.e. terms inside sums in the expression (31) do not depends on $k$, this can be simplified:

$$\begin{aligned}I^{SE}(2\tau;\tau) &= 1 - \frac{1}{N_s}\sum_k \exp\left\{-\frac{1}{2}\sum_m \left\langle\tilde{\varphi}_{km}^d(\tau)\cdot\varphi_{km}^d(\tau)\right\rangle\right\} = \\ &= 1 - \exp\left\{-\frac{1}{2N_s}\sum_{k;m}\left\langle\tilde{\varphi}_{km}^d(\tau)\cdot\varphi_{km}^d(\tau)\right\rangle\right\}\end{aligned} \tag{32}$$

### 3. Discussion.

Expressions (20), (21), (22), (22), (25), (26), (27), (31) and (32) represent the main theoretical results of this paper. They resolve, in a very general form, the problem of calculating two types of solid echo signals $S_1$, $S_2$ and the construction of the $I^{SE}(2\tau;\tau)$ function for many-spin systems, when the Hamiltonian is described by expression (1). The main approximation which was made at course of derivation is the modified Anderson-Weiss approximation firstly formulated in [46]. It is exact for times $t \ll T_2$ and takes into account, at variance with the ordinary Anderson-Weiss approximation, flip-flop processes causing spin-diffusion for times $t > T_2$. It is clearly seen from all the discussed expressions that all of them contain both intra- and intermolecular magnetic dipole-dipole interactions between proton spins.

It is instructive to compare expression (32) for the $I^{SE}(2\tau;\tau)$ function with the so-called DQ build up function $I_{nDQ}(\tau_{DQ})$ measured by double quantum resonance, which for many-spins systems was firstly derived in [47]:

$$I_{nDQ}(\tau_{DQ}) = \frac{1}{2}\left(1 - \exp\left\{-\frac{2}{N_s}\sum_{k,m}\langle\varphi_{km}^{ex}\varphi_{km}^{rec}\rangle\right\}\right), \qquad (33)$$

where:

$$\varphi_{km}^{ex} = \frac{\gamma_H^2 \hbar}{2}\int_0^{\tau_{DQ}}\frac{3\cos^2\theta_{km}(t_1) - 1}{r_{km}^3(t_1)}dt_1$$

$$\varphi_{km}^{rec} = \frac{\gamma_H^2 \hbar}{2}\int_{\tau_{DQ}}^{2\tau_{DQ}}\frac{3\cos^2\theta_{km}(t_1) - 1}{r_{km}^3(t_1)}dt_1 \qquad (34)$$

and $\tau_{DQ}$ is the duration of the excitation and reconversion periods, which are experimentally controlled parameters, i.e. the DQ evolution time.

By comparing expressions (21), (22), (32) with (33) and (34) it is possible to identify the close similarity between the $I^{SE}(2\tau;\tau)$ which therefore is convenient to name, in analogy with DQ resonance, the proton solid echo build-up function measured by the superposition of two kinds of solid echo and Hahn echo, i.e. signals $S_1$, $S_2$ and $S_3$, and proton DQ build up function $I_{nDQ}(\tau_{DQ})$ measured by DQ resonance. Indeed the interval between two echoes $\tau$ is the analogue of the duration of the excitation and reconversion periods $\tau_{DQ}$, and the quantities $\varphi_{kl}^d(\tau)$ and $\tilde{\varphi}_{km}^d$ are analogues of $\varphi_{km}^{ex}$ and $\varphi_{km}^{rec}$. A small formal difference is connected with $P_{kl}^{fl}(t_1)$, i.e. the probability for a given pair of spins with numbers $k$ and $l$ not to participate in flip-flop processes in the time interval $t_1$. This difference requires various approximations, which were used in [47] at course of derivation of the expressions (33) and (34) and in this paper. Actually, in the cited paper we did not use the modified Anderson-Weiss approximation, but the assumptions which are equivalent to the ordinary Anderson-Weiss approximation. This is why in expressions (34) the probability $P_{kl}^{fl}(t_1)$ is absent, i.e. it is assumed to be equal to unity. However, without difficulties modified Anderson-Weiss approximation can be introduced also for the derivation of $I_{nDQ}(\tau_{DQ})$ and expressions (34) will look like:

$$\varphi_{km}^{ex} = \frac{\gamma_H^2 \hbar}{2} \int_0^{\tau_{DQ}} \frac{3\cos^2\theta_{km}(t_1)-1}{r_{km}^3(t_1)} P_{kl}^{fl}(t_1) dt_1$$

$$\varphi_{km}^{rec} = \frac{\gamma_H^2 \hbar}{2} \int_{\tau_{DQ}}^{2\tau_{DQ}} \frac{3\cos^2\theta_{km}(t_1)-1}{r_{km}^3(t_1)} P_{kl}^{fl}(t_1) dt_1$$

(35)

Note that these discussed probabilities are important for long times $t > \frac{1}{2}\tau_{fl}$, where $\tau_{fl}$ is the characteristic time for flip-flop transitions for individual spin (the factor ½ reflects the fact that the probability $P_{kl}^{fl}(t_1)$ not to have a flip-flop transition is related to both spins simultaneously.

From results of [46], see expressions (53) and (54) therein, it follows that $\tau_{fl} \approx 8^{\frac{2}{4-3\alpha}} T_2 > \sqrt{8} T_2$, where $\frac{1}{4} < \alpha < \frac{2}{3}$ is the power of the polymer segments mean squared displacement time dependence exponent $\langle r_n^2(t) \rangle \propto t^\alpha$, and $\tau_{fl} \approx 6T_2$ for $\alpha > \frac{2}{3}$. The long time behavior experimentally observed in $I^{SE}(2\tau;\tau)$ or $I_{nDQ}(\tau_{DQ})$ is actually very difficult for precise theoretical interpretations, because they are essentially determined by many-spins correlations, i.e. with unresolved theoretical problem. Therefore at this point we are forced to discuss only different approximations. This is not a point of this paper and we restrict ourselves only to general qualitative remarks.

Let us assume a polymer melt with very large molecular mass, so that the terminal relaxation time $\tau_1 \gg T_2$. One can then discuss differences in the limiting behavior of functions $I^{SE}(2\tau;\tau)$ and $I_{nDQ}(\tau_{DQ})$ at very large times, i.e. $\tau, \tau_{DQ} \gg T_2$ when the calculation is based on either the ordinary or the modified Anderson-Weiss approximation. In the case of the ordinary Anderson-Weiss approximation, effects of spin-diffusion are absent, then $I^{SE;AW}(\infty;\infty)=1$ and $I_{nDQ}^{AW}(\infty)=\frac{1}{2}$ because $\frac{1}{2N_s}\sum_{k;m}\langle\tilde{\varphi}_{km}^{d}(\tau)\cdot\varphi_{km}^{d}(\tau)\rangle\xrightarrow{\tau\to\infty}\infty$ and $\frac{2}{N_s}\sum_{k,m}\langle\varphi_{km}^{ex}\varphi_{km}^{rec}\rangle\xrightarrow{\tau_{DQ}\to\infty}\infty$. In the case of the modified Anderson-Weiss approximation, spin-diffusion destroys correlations between spin rotations at times $\tau,\tau_{DQ}\gg\tau_{fl}$ and therefore the mentioned sums will have finite limits. To illustrate this it is convenient to consider the case of a rigid lattice, i.e. in a situation when one can neglect motions of protons. In this case, as can be seen from (18),

$$P_{kl}^{fl}(t)\approx\exp\left\{-\frac{M_2}{8}t^2\right\}, \tag{36}$$

where $M_2=\frac{9}{16}\frac{1}{\hbar^2}\frac{1}{N_s}\sum_{k,l}'\langle A_{kl}^2(0)\rangle$ is the second moment of proton spins in the rigid lattice, including both intermolecular and intramolecular magnetic dipole-dipole interactions, while considering the case $\tilde{A}_{kl}=A_{kl}$.

Using relations (32), (21), (22) and the normal form representation (see remarks after the expression (17)) the function $I^{SE}(2\tau;\tau)$ can then be written as follows:

$$I^{SE}(2\tau;\tau)=1-\exp\left\{-\frac{1}{2}M_2\int_0^\tau dt_1(\tau-t_1)\left[\exp\left\{-\frac{M_2}{8}(\tau+t_1)^2\right\}+\exp\left\{-\frac{M_2}{8}(\tau-t_1)^2\right\}\right]\right\}=$$
$$1-\exp\left\{-2\left(1+\exp\left\{-\frac{M_2\tau^2}{2}\right\}-2\exp\left\{-\frac{M_2\tau^2}{8}\right\}+\sqrt{\frac{\pi}{2}}M_2^{1/2}\tau\left[erf\left(\frac{M_2^{1/2}\tau}{\sqrt{2}}\right)-erf\left(\frac{M_2^{1/2}\tau}{2\sqrt{2}}\right)\right]\right)\right\}, \tag{37}$$

where $erf(x)=\frac{2}{\sqrt{\pi}}\int_0^x\exp\{-t^2\}dt$.

For a model rigid lattice this long time limit can be calculated exactly, and the results are as follows: $I^{SE;mAW}(\infty;\infty)=1-\frac{1}{e^2}$ and $I_{nDQ}^{mAW}(\infty)=\frac{1}{2}\left(1-\frac{1}{e^2}\right)$. One can see that changes are relatively small, but nevertheless observable. This is because the characteristic flip-flop time is considerably longer than the effective spin-spin relaxation time for rigid lattice $\tau_{fl}\approx\sqrt{8}T_2$. In Figure 1 one can see the difference between the ordinary and the modified Anderson-Weiss approximations for the function $I^{SE}(2\tau;\tau)$, the second moment $M_2 = 2$, $T_2 \approx \sqrt{\frac{2}{M_2}} = 1$, this means that time is measured in unites of $T_2$ for ordinary Anderson-Weiss approximation.

Note also, that the long time limit will exist only for the situations, when a molecular mass distribution is not very large. If for example a system under study contains fast low molecular fractions, then after constriction the nominator of the expression (28) does not contain their contributions, but denominator does and at long enough time this contribution will become dominant. Therefore the solid echo build up function $I^{SE}(2\tau;\tau)$ for the discussed situation will decrease with time after reaching its maximum. This, however, will be discussed in our future paper.

Now let us discuss the behavior of the mentioned functions at the initial moments of time in more details: when $\tau, \tau_{DQ} \ll T_2$ and $P_{km}^{fl}(\tau) \cong 1$, one can neglect flip-flop processes and the arguments of exponents are small. Decomposing expressions (32) and (33) to Taylor series and keeping only terms quadratic in phases one obtains:

$$I^{SE}(2\tau;\tau) = \frac{1}{2N_s} \sum_{k;m} \langle \tilde{\varphi}_{km}^d(\tau) \cdot \varphi_{km}^d(\tau) \rangle + ...$$

$$I_{nDQ}(\tau_{DQ}) = \frac{1}{N_s} \sum_{k,m} \langle \varphi_{km}^{ex} \varphi_{km}^{rec} \rangle + ...$$
(38)

Then using expressions (21), (22), (32) and employing the translational invariance of the time dependent correlation functions, (38) can be rewritten as the following:

$$I^{SE}(2\tau;\tau) = \frac{9}{8}\gamma^4\hbar^2 \int_0^\tau (\tau - t_1)\{A_0^d(\tau + t_1) + A_0^d(\tau - t_1)\} dt_1 + ...$$

$$I_{nDQ}(\tau_{DQ}) = \frac{1}{4}\gamma^4\hbar^2 \int_0^{\tau_{DQ}} (\tau_{DQ} - t_1)\{A_0^d(\tau_{DQ} + t_1) + A_0^d(\tau - t_1)\} dt_1 + ...$$
(39)

where

$$A_0^d(t) = \frac{1}{N_s} \sum_{k,m} \left\langle \frac{1 - 3\cos(\theta_{km}(t))}{r_{km}^3(t)} \cdot \frac{1 - 3\cos(\theta_{km}(0))}{r_{km}^3(0)} \right\rangle.$$
(40)

One can see that the initial rise of both functions differ from each other only by a numerical coefficient and at the following we will discuss only $I^{SE}(2\tau;\tau)$ for the sake of brevity.

The time dependent total dipole-dipole correlation function $A_0^d(t)$ for polymer melts was analyzed in detail earlier [47]. It can be separated into a sum of intermolecular and intramolecular parts, corresponding to the contributions from protons from different and same macromolecules, respectively:

$$A_0^d(t) = A_0^{d;\text{inter}}(t) + A_0^{d;\text{intra}}(t). \tag{41}$$

The intermolecular contribution, for times much longer than the segmental relaxation time $t \gg \tau_s$ is related to the relative mean squared displacements of proton spins from different macromolecules $\langle \tilde{r}^2(t) \rangle$ by the following expression:

$$A_0^{d;\text{inter}}(t) = \sqrt{\frac{2}{3\pi}} \frac{16\pi}{5} \frac{n_s}{\langle \tilde{r}^2(t) \rangle^{3/2}}, \tag{42}$$

where $n_s$ is the concentration of protons in the system.

Correspondingly to (39), the experimentally measurable function $I^{SE}(2\tau;\tau)$ can also be represented as the sum of intramolecular and intermolecular contributions:

$$I^{SE}(2\tau;\tau) = I^{SE;\text{intra}}(2\tau;\tau) + I^{SE;\text{inter}}(2\tau;\tau). \tag{43}$$

Using deuteration technique as was carried out in [17] for the spin-lattice relaxation dispersion, it would be possible to also separate the intermolecular contribution from the intramolecular contribution in spin-echo experiments according to (41). Using relation (40) one then finds that the intermolecular part is connected in a rather simple way with the relative mean-squared displacements of polymer segments from different macromolecules:

$$I^{SE;\text{inter}}(2\tau;\tau) = \frac{18\pi}{5}\sqrt{\frac{2}{3\pi}}\gamma^4\hbar^2 n_s \int_0^\tau (\tau-t_1)\left\{\frac{1}{\langle\tilde{r}^2(\tau+t_1)\rangle^{3/2}} + \frac{1}{\langle\tilde{r}^2(\tau-t_1)\rangle^{3/2}}\right\}dt_1 + \ldots. \tag{44}$$

For those cases when the relative mean-squared displacements of polymer segments from different macromolecules can be described by a time independent exponent $\alpha$, i.e. $\langle\tilde{r}^2(t)\rangle = A \cdot t^\alpha$, integration of the right-hand side of expression (44) can be performed exactly, where for integration it is sufficient to assume $\alpha < 2$:

$$I^{SE;inter}(2\tau;\tau) = \frac{18\pi}{5} f(\alpha) \sqrt{\frac{2}{3\pi}} \gamma^4 \hbar^2 n_s \frac{\tau^2}{\langle \tilde{r}^2(\tau) \rangle^{3/2}} + ..., \tag{45}$$

where

$$f(\alpha) = \frac{2^{1-\frac{3\alpha}{2}} - 1}{\left(1 - \frac{3\alpha}{2}\right)\left(1 - \frac{2\alpha}{4}\right)} \tag{46}$$

is a numerical coefficient.

From equation (45) one can obtain the relative mean-squared displacement of polymer segments from different polymer chains:

$$\langle \tilde{r}^2(\tau) \rangle = \left( \frac{18\pi}{5} f(\alpha) \sqrt{\frac{2}{3\pi}} \gamma^4 \hbar^2 n_s \frac{\tau^2}{I^{SE;inter}(2\tau;\tau)} \right)^{\frac{2}{3}}. \tag{47}$$

Note that the characteristic flip-flop time is considerably longer that the spin-spin relaxation time $\tau_{fl} \approx 8^{\frac{2}{4-3\alpha}} T_2 > \sqrt{8} T_2$, as was already mentioned. Then for times $\tau < \tau_{fl} \approx 8^{\frac{2}{4-3\alpha}} T_2$ expression (32) can be rewritten as:

$$I^{SE;inter}(2\tau;\tau) =$$
$$= 1 - \exp\left\{ -\frac{18\pi}{5} \sqrt{\frac{2}{3\pi}} \gamma^4 \hbar^2 n_s \int_0^\tau (\tau - t_1) \left\{ \frac{1}{\langle \tilde{r}^2(\tau+t_1) \rangle^{3/2}} + \frac{1}{\langle \tilde{r}^2(\tau-t_1) \rangle^{3/2}} \right\} dt_1 + ... \right\}. \tag{48}$$

The relative mean-squared displacement of polymer segments from different macromolecules for time-independent $\alpha$ can then be calculated from the following relation:

$$\langle \tilde{r}^2(\tau) \rangle = \left( \frac{18\pi}{5} f(\alpha) \sqrt{\frac{2}{3\pi}} \gamma^4 \hbar^2 n_s \frac{\tau^2}{\ln\left(\frac{1}{1 - I^{SE;inter}(2\tau;\tau)}\right)} \right)^{\frac{2}{3}}. \tag{49}$$

If the Anderson-Weiss approximation was exact, then (49) would also be exact. The modified Anderson-Weiss approximation begins to differ from the standard one when $\tau > \frac{1}{2}\tau_{fl} > T_2$ Therefore it would be a reasonable approach to test the validity of

(49) by real experiments within the domain $\tau < \frac{1}{2}\tau_{fl}$, which is larger than that for expression (47), which is applicable when $\tau < T_2$.

In conclusion it can be said that a systematic investigation of intermolecular magnetic dipole-dipole contributions to proton signal of differently encoded echoes in polymer melts can give important experimental information about relative mean squared displacement of polymer segments from different macromolecules on a millisecond time scale. Corresponding experiments are in progress.

**Acknowledgments.**

Financial support from Deutsche Forschungsgemeinschaft (DFG) through grant STA 511/13-1 is gratefully acknowledged.

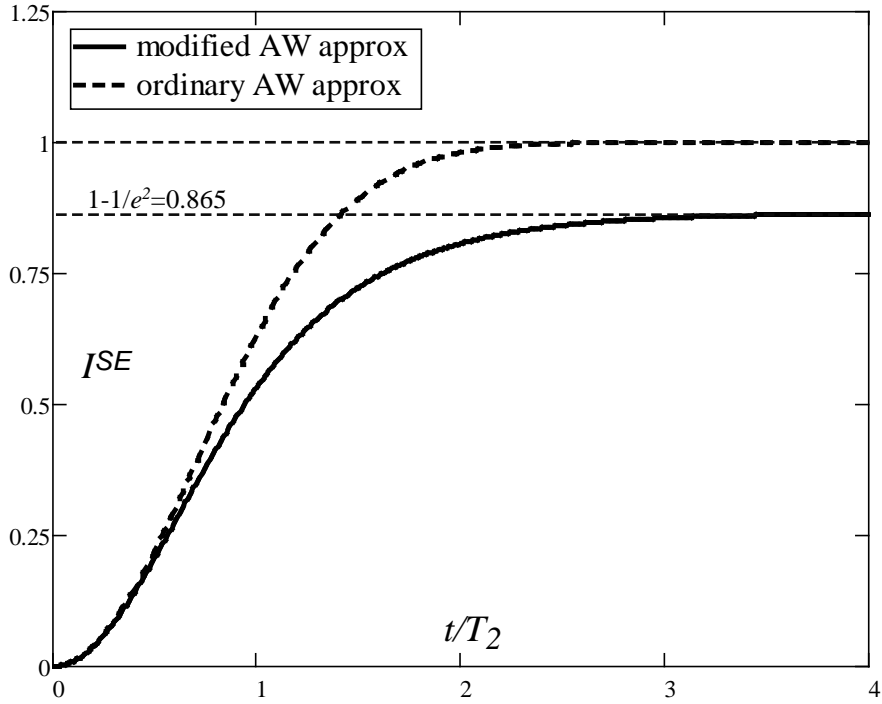

FIG. 1. Time dependence of the solid echo build up function $I^{SE}(2\tau;\tau)$ for the case of immobile lattice: in the case of ordinary Anderson-Weiss approximation (--) and in the case of modified Anderson-Weiss approximation (−). X-axis is represented in units of $\frac{t}{T_2}$, where $T_2 = \sqrt{\frac{2}{M_2}}$ is the effective spin-spin relaxation time for the case of ordinary Anderson-Weiss approximation.